\documentclass[%
reprint,
superscriptaddress,
showpacs,
 amsmath,amssymb,
 aps,
prb,
floatfix,
]{revtex4-1}

\usepackage[]{subfigure}
\usepackage{graphicx}
\usepackage{dcolumn}
\usepackage{bm}

\begin{document}

\preprint{APS/123-QED}

\title{Magnetometry with nitrogen-vacancy ensembles in diamond based on infrared absorption in a doubly resonant optical cavity}
\author{Y.~Dumeige}
\email{yannick.dumeige@univ-rennes1.fr}
\affiliation{UEB, Universit\'e Europ\'eenne de Bretagne, Universit\'e de Rennes I and CNRS, UMR 6082 FOTON, Enssat, 6 rue de Kerampont, CS 80518, 22305 Lannion cedex, France}
\author{M.~Chipaux}
\affiliation{Thales Research and Technology, Campus Polytechnique, 91767 Palaiseau, France}%
\author{V.~Jacques}
\affiliation{Laboratoire Aim\'e Cotton, CNRS, Universit\'e Paris-Sud and ENS Cachan, 91405 Orsay, France}%
\author{F.~Treussart}
\affiliation{Laboratoire Aim\'e Cotton, CNRS, Universit\'e Paris-Sud and ENS Cachan, 91405 Orsay, France}%
\author{J.-F.~Roch}
\affiliation{Laboratoire Aim\'e Cotton, CNRS, Universit\'e Paris-Sud and ENS Cachan, 91405 Orsay, France}%
\author{T.~Debuisschert}
\affiliation{Thales Research and Technology, Campus Polytechnique, 91767 Palaiseau, France}%
\author{V.~Acosta}
\affiliation{Hewlett-Packard Laboratories, Palo Alto, CA, USA}
\affiliation{Department of Physics, University of California at Berkeley, CA 94720-7300 USA}%
\author{A.~Jarmola}
\author{K.~Jensen}
\author{P.~Kehayias}
\author{D.~Budker}
\affiliation{Department of Physics, University of California at Berkeley, CA 94720-7300 USA}%

\date{\today}

\begin{abstract}
We propose to use an optical cavity to enhance the sensitivity of magnetometers relying on the detection of the spin state of high-density nitrogen-vacancy
ensembles in diamond using infrared optical absorption. The role of the cavity is to obtain a contrast in the absorption-detected magnetic resonance approaching unity at room temperature. We project an increase in the photon shot-noise limited sensitivity of two orders of magnitude in comparison with a single-pass approach. Optical losses can limit the enhancement to one order of magnitude which could still enable room temperature operation. Finally, the optical cavity also allows to use smaller pumping power when it is designed to be resonant at both the pump and the signal wavelength. 
\end{abstract}

\maketitle


\section{Introduction}

The negatively charged nitrogen-vacancy (NV$^-$) center in diamond can be used as a solid-state magnetic sensor due to its electron spin resonance (ESR). The center can be optically polarized and its polarization detected through the spin-state dependence of the luminescence \cite{Jelezko06, Taylor08}. Sensors based on a single NV$^-$ center have the potential to achieve atomic-scale spatial resolution \cite{Degen08, Maze08, Balasubramanian08}. On the other hand, magnetic field sensitivity can be enhanced by engineering the diamond material in order to increase the spin dephasing time which limits the ESR linewidth \cite{Balasubramanian09}. The magnetic response of an ensemble of NV$^-$ centers \cite{Lai09, Maertz10, Steinert10, Pham11} leads to a luminescence magnified by the number $N$ of the sensing spins. Such collective response also improves the signal to noise ratio and the sensitivity by a factor $\sqrt{N}$ since the quantum projection noise associated with the spin-state determination scales as \cite{Itano93, Taylor08} $\sqrt{N}$.

Currently, the sensitivity of practical magnetometers based on the detection of red luminescence of the NV$^-$ ensemble is limited by background fluorescence and poor collection efficiency. Recent advances in diamond engineering have enabled improvements in collection efficiency which should improve fluorescence based sensors,\cite{Hadden10, Siyushev10, Marseglia11, LeSage2012, Maletinsky2012} but here we consider a different approach. In addition to the well-known transitions leading to red fluorescence, it has been shown recently the existence of an infrared (IR) transition related to the singlet states\cite{Rogers08, Acosta10prb}. This transition can be exploited in an IR-absorption scheme with an increased sensitivity as compared to the usual scheme\cite{Acosta10}. In this paper we show that using IR absorption detection in combination with a high-finesse optical cavity, it is possible to tune the absorption contrast to order unity thereby dramatically improving the magnetic field sensitivity. We first recall the parameters which set the magnetometer sensitivity. We then theoretically investigate the extension of this detection scheme to the case where the diamond crystal hosting the NV$^-$ ensemble is inserted inside a high-finesse optical cavity, as it is usually done in cavity ring-down spectroscopy \cite{Berden00}. Finally we determine the improvement of the magnetometer response associated with the cavity quality (Q) factor.\\

\section{Single-pass photon shot-noise limited magnetic field sensitivity}

\begin{figure}[h!]
\includegraphics[width=8cm]{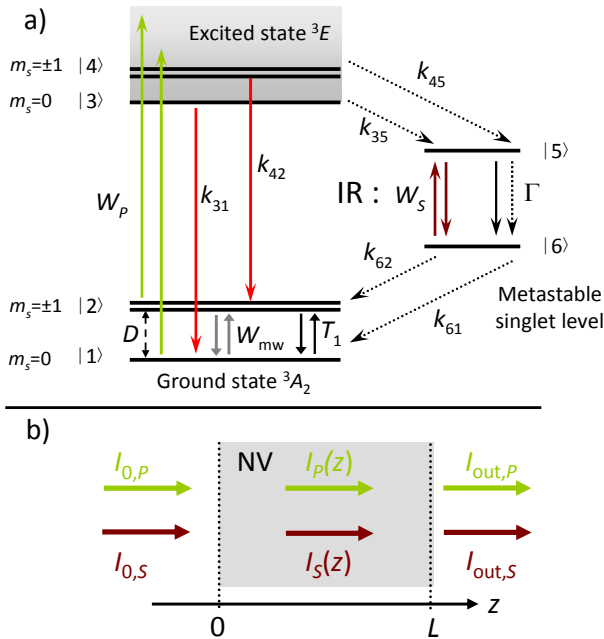}
\caption{\label{level} a) Level structure of NV$^-$ center in diamond. The photophysical parameters related to this six-level system are given in Tab. \ref{tab_nv} of Appendix \ref{tableaux}. The solid (dot) lines correspond to radiative (non-radiative) transitions. $D\approx2.87~\mathrm{GHz}$ is the zero-field splitting of the ground state. b) Diagram of the experimental configuration used to measure the single-pass contrast of the IR absorption under resonant microwave application\cite{Acosta10}. $I_{0,P}$ and $I_{0,S}$ are the pump (wavelength $\lambda_P$) and the probe input intensities.}
\end{figure}

The principle of the method is similar to the one used in optical magnetometers based on the precession of spin-polarized atomic gases\cite{Budker07}. The applied magnetic field value is obtained by optically measuring the Zeeman shifts of the NV$^-$ defect spin sublevels via the absorption monitoring of the IR probe signal. The photodynamics of NV$^-$ centers are modeled using the level structure depicted in Fig \ref{level}.a). The spin sublevels $m_s=0$ and $m_s=\pm1$ of the $^3A_2$ ground triplet state are labeled $\left|1\right\rangle$ and $\left|2\right\rangle$ and separated by $D=2.87~\mathrm{GHz}$ in zero magnetic field. $\left|3\right\rangle$ and $\left|4\right\rangle$ are the respective spin sublevels of the $^3E$ excited level. Levels $\left|5\right\rangle$ and $\left|6\right\rangle$ are single-state levels related to the infrared absorption transition. The relaxation rate from state $i$ to $j$ is denoted $k_{ij}$. As $k_{35}\ll k_{45}$ (see Table \ref{tab_nv} in Appendix \ref{tableaux}), the system is optically polarized in $m_s=0$ while pumping the NV$^-$ centers via the phonon sideband. Without microwaves applied, there is reduced population in the metastable singlet state, $\left|6\right\rangle$, corresponding to a minimal IR absorption signal. Under application of resonant microwaves with frequency $D\pm\gamma B/(2\pi)$, where $B$ is the magnetic field projection along one of the four NV$^-$ orientations and $\gamma=1.761\times10^{11}~\mathrm{s}^{-1}\mathrm{T}^{-1}$ is the gyromagnetic ratio, population is transfered from $m_s=0$ to $m_s=\pm 1$ sublevel resulting in greater population in the metastable singlet and lower IR signal transmission. The experimental configuration for single-pass absorption measurements is shown in Fig \ref{level}.b).
The output transmission is measured either with or without applying the resonant microwaves. The contrast $\mathcal{C}$ is defined as the relative difference in the IR signal detected after propagation in the diamond crystal of length $L$
\begin{equation}\label{Contrast_single_pass}
\mathcal{C}=\frac{I_{\mathrm{out},S}(0)-I_{\mathrm{out},S}(\Omega_R)}{I_{\mathrm{out},S}(0)},
\end{equation}
where $I_{\mathrm{out},S}(0)$ [$I_{\mathrm{out},S}(\Omega_R)$] denotes the IR signal intensity without [with] the application of the microwave field whose Rabi angular frequency is denoted $\Omega_R$. We can estimate the photon shot-noise limited sensitivity at room temperature for an optical power compatible with the IR saturation intensity. For an ESR full-width-at-half-maximum (FWHM) $\Gamma_{\mathrm{mw}}$, the magnetic field sensitivity (or the minimum detectable magnetic field) of a magnetometer based on IR absorption measurement is given by \cite{Acosta10,Dreau11, Rondin12}
\begin{equation}\label{sensitivity}
\delta B=\frac{\Gamma_{\mathrm{mw}}}{\gamma\mathcal{C}}\sqrt{\frac{hc}{P_{S}t_{\mathrm{m}}\lambda_S}},
\end{equation}
where $P_S$ is the measured IR probe beam signal output power (wavelength $\lambda_S$), and $t_{\mathrm{m}}$ is the measurement time. Assuming no power broadening from either pump or microwaves, the ESR FWHM is related to the electron spin dephasing time by $\Gamma_{\mathrm{mw}}=2/T_2^*$ (in $\mathrm{rad/s}$). For a detected IR signal power $P_{S}=300~\mathrm{mW}$ using Eq. (\ref{sensitivity}) with parameter values given in Tab. \ref{tab_nv2} of Appendix \ref{tableaux} we obtain a shot-noise limited magnetic field sensitivity of $20~\mathrm{pT}/\sqrt{\mathrm{Hz}}$ in a single-pass configuration at room temperature. Note that considering this IR signal power and a beam waist diameter of $2w_0=50~\mu\mathrm{m}$ there is no saturation of the IR absorption (see Appendix \ref{cross_section_estimation}). For this single-pass configuration, the contrast cannot be improved by increasing the thickness of the sample since for $L$ larger than the pump penetration depth ($\approx120~\mathrm{\mu m}$ from the absorption cross section and NV$^-$ center density of Tab. \ref{tab_nv} and \ref{tab_nv2}) its absorption becomes too strong. The photon shot-noise limited sensitivity can be compared to the spin-noise limited sensitivity
\begin{equation}\label{sql}
\delta B_{\mathrm{q}}=\frac{2}{\gamma\sqrt{n V T_2^*t_\mathrm{m}}},
\end{equation}
where we take into account through the factor of $2$ that only one fourth of the NV$^-$ centers are oriented along the magnetic field\cite{Pham12}, $n$ is the NV$^-$-center density and $V$ is the illuminated diamond volume. In the single pass configuration of Ref. [\onlinecite{Acosta10}], the spin-noise limited sensitivity is about $0.02~\mathrm{pT}/\sqrt{\mathrm{Hz}}$.

\section{Sensitivity enhancement}

According to Eq. (\ref{sensitivity}), the magnetic-field sensitivity is limited by the low contrast $\mathcal{C}$. In particular, at room temperature the contrast is an order of magnitude smaller than at $75~\mathrm{K}$ due to homogeneous broadening \cite{Acosta10}. It can also be seen as limited by the optical depth estimated to only $2.2\times10^{-2}$ for the experimental demonstration reported in Ref. [\onlinecite{Acosta10}]. However, the optical depth can be increased by using a cavity resonant at the IR signal wavelength resulting in an increase of the optical path by a factor proportional to the finesse of the cavity. Moreover, using a diamond crystal thickness smaller than the pump absorption length allows to overcome the issue of the pump depletion and to obtain a good microwave field homogeneity along the crystal. 
\begin{figure}[h!]
\includegraphics[width=8.5cm]{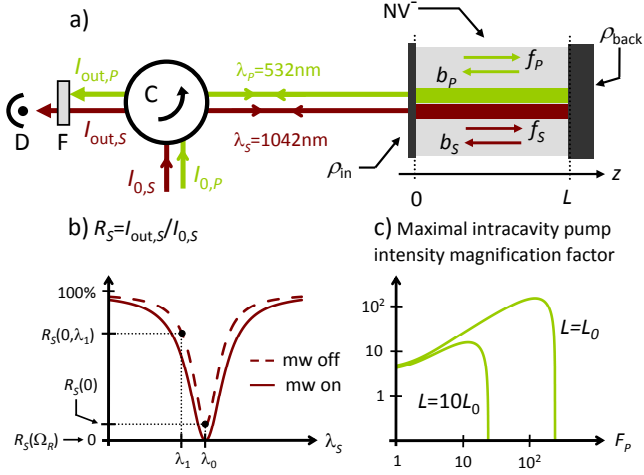}
\caption{\label{cavite} a) All pass cavity (we consider a perfectly reflecting backside mirror $\left|\rho_{\mathrm{back},S}\right|=1$) used for magnetic field sensitivity enhancement. $\rho_{\mathrm{in,i}}$ is the amplitude reflectivity of the input coated mirror. The cavity can be doubly resonant for the pump and the signal. C: optical circulator, F: optical filter rejecting the pump beam, D: optical detector. b) Reflected spectrum from the cavity for switched-on or switched-off microwaves (mw) resonant at the level $\left|1\right\rangle$-$\left|2\right\rangle$ transition. $\lambda_0$ is the IR cavity resonance wavelength. c) Maximal intracavity pump beam optical power magnification factor for a given value of intracavity absorption and two values of cavity lengths $L=L_0$ and $L=10L_0$. The finesse of the cavity at the pump wavelength is denoted $F_P$. $L_0$ is the cavity length which gives a critical coupling (and thus the the optimal magnification factor) for $F_P\approx100$. Note that even with $\rho_{\mathrm{in},P}=0$, the all-pass configuration gives a maximal magnification around $4$ due to reflection on the backside mirror.}
\end{figure}
We consider the Fabry-Perot cavity configuration depicted in Fig \ref{cavite}a), consisting of a two-side coated bulk-diamond plate containing a high NV$^-$-center density (larger than $4\times10^{23}~\mathrm{m}^{-3}$). We consider an all-pass Fabry-Perot cavity for the IR signal. This means that the amplitude reflectivity of the back mirror is $\rho_{\mathrm{back},S}=1$ and of the input mirror reflectivity is $\rho_{\mathrm{in},S}<1$. Regarding the pump, we consider either single-pass propagation ($\rho_{\mathrm{in},P}=\rho_{\mathrm{back},P}=0$) or all-pass cavities ($\rho_{\mathrm{back},P}=1$). We define the reflection of the cavity at optical resonance by $R_{i}=I_{\mathrm{out,i}}/I_{0,i}$ with (with $i\in\{P,S\}$).

\subsection{Basic principle of the cavity effect}

The complete analysis of the cavity has to be performed numerically. In order to allow a simple interpretation of the results, we first derive analytical expressions for the sensitivity assuming no saturation of the IR-signal absorption. The absorption of the IR signal due to levels $\left|5\right\rangle$ and $\left|6\right\rangle$ and the spin polarization due to the pump beam system is simply taken into account by $A_S$ the single-pass round-trip amplitude transmission. We also assume a good finesse cavity at the IR signal wavelength and thus the input mirror reflectivity can be written $\rho_{\mathrm{in},S}=1-\varepsilon$ with $\varepsilon\ll1$. With the application of the resonant microwave field we have: $A_S(\Omega_R)=1-a_{\Omega_R}$ ($a_{\Omega_R}\ll1$) whereas for an off-resonance microwave field we have: $A_S(0)=1-a_{0}$ ($a_{0}\ll1$). We define the optically resonant reflectivity for respectively off- and on-resonance microwave fields using the results given in Appendix \ref{analy} at the first order
\begin{equation}
  \left\{
    \begin{aligned}
      R_S(0) & = \left(\frac{\varepsilon-a_{0}}{\varepsilon+a_{0}}\right)^2\\
			R_S(\Omega_R) & = \left(\frac{\varepsilon-a_{\Omega_R}}{\varepsilon+a_{\Omega_R}}\right)^2.\\
    \end{aligned}
  \right.
\end{equation}
The finesse of the cavity given in Eq. (\ref{finesse_appendix}) can also be written at the first order in $\varepsilon$ and $a_i$ 
\begin{equation}\label{good_finesse}
F_S=\frac{\pi}{\varepsilon+a_i}.
\end{equation}
where $i=0$ for off-resonance microwaves and $i=\Omega_R$ for on-resonance microwaves.

\subsubsection{Optimal cavity coupling}

Assuming a perfect spin polarization and no additional optical losses, we have: $a_0=0$. In this case, $R_S(0)=1$ and thus the off-resonance reflected detected signal is equal to the input signal power $P_{0,S}$. The contrast reads $\mathcal{C}=1-R_S(\Omega_R)$ and the magnetic field sensitivity is given by
\begin{equation}\label{sensitivity_sans_perte}
\delta B=\frac{\Gamma_{\mathrm{mw}}}{\gamma\left[1-R_S(\Omega_R)\right]}\sqrt{\frac{hc}{P_{0,S}t_{\mathrm{m}}\lambda_S}}.
\end{equation}
For $\varepsilon=a_{\Omega_R}$, the incoming and outgoing fields destructively interfere at the resonant wavelength and $R_{S}(\Omega_R)=0$. The laser probe beam is then critically coupled \cite{Dumeige08} to the cavity-NV$^-$ ensemble system and the contrast is equal to $1$. For this particular value the optimal sensibility of the magnetometer is reached.

\subsubsection{Effects of the microwave off-resonance absorption}

Now we consider the more realistic case of a non ideal spin-polarization and material with parasitic IR losses which gives $a_{\Omega_R}>a_0>0$. There are three possible cases
\begin{equation}\label{cas_possible}
  \left\{
    \begin{aligned}
      \mathrm{i)}~~\varepsilon>\sqrt{a_{0}a_{\Omega_R}} & ~~~R_S(0)>R_S(\Omega_R)\\
      \mathrm{ii)}~~\varepsilon=\sqrt{a_{0}a_{\Omega_R}} & ~~~R_S(0)=R_S(\Omega_R)\\
			\mathrm{iii)}~~\varepsilon<\sqrt{a_{0}a_{\Omega_R}} & ~~~R_S(0)<R_S(\Omega_R).\\
    \end{aligned}
  \right.
\end{equation}
Consequently, depending on the relative value of $R_S(0)$ and $R_S(\Omega_R)$, the expression of the contrast is different. This can be taken into account by writing
\begin{equation}
\mathcal{C}=\frac{\left|R_S(0)-R_S(\Omega_R)\right|}{\max\left[R_S(\Omega_R),R_S(0)\right]}.\label{contrast_cav}
\end{equation}
This relation can be used to write the expression of the minimum detectable magnetic field taking into account the detrimental effect of the residual IR absorption due to non-ideal branching ratio to the metastable state by multiplying $P_{0,S}$ by $\max\left[R_S(\Omega_R),R_S(0)\right]$ to obtain the detected IR power $P_S$ of Eq. (\ref{sensitivity}). The fundamental advantage of the present method is that this quantity falls under the square root whereas for methods based on the visible-fluorescence monitoring the non-ideal branching ratio reduces the contrast $\mathcal{C}$ by a similar amount, but this quantity falls outside the square root. One can estimate that in the same conditions, the minimal detectable magnetic field $\delta B$ is reduced by a factor of $\approx5$ in comparison with $\delta B_{\mathrm{f}}$ obtained via fluorescence method with a collection efficiency $\eta\approx0.47$ (see details and discussion in Appendix \ref{fundamental}). The sensitivity thus reads
\begin{equation}\label{sensitivity_cav}
\delta B=\frac{\Gamma_{\mathrm{mw}}}{\gamma\left|R_S(0)-R_S(\Omega_R)\right|}\sqrt{\frac{hc\times\max\left[R_S(\Omega_R),R_S(0)\right]}{P_{0,S}t_{\mathrm{m}}\lambda_S}}.
\end{equation}
In the present case, there are two critical-coupling conditions, thus the sensitivity $\delta B$ can reach two optimal values obtained for $\varepsilon=a_{\Omega_R}$ (solid line in Fig \ref{cavite}b) or $\varepsilon=a_{0}$. Note that due to the factor $\sqrt{\max\left[R_S(\Omega_R),R_S(0)\right]}$ in the numerator of Eq. (\ref{sensitivity_cav}), the minimum values of $\delta B$ is actually reached for values of $\varepsilon$ slightly different from the exact critical-coupling finesse. This will be accurately described in the numerical calculations. 
We first consider the case i) of Eqs. (\ref{cas_possible}). Assuming $\varepsilon\gg a_{\Omega_R}$ we have
\begin{equation}\label{sensitivity_low_finesse}
\delta B\approx\frac{\pi\Gamma_{\mathrm{mw}}}{4\gamma F_S(a_{\Omega_R}-a_0)}\sqrt{\frac{hc}{P_{0,S}t_{\mathrm{m}}\lambda_S}}.
\end{equation}
This means that for low cavity finesses the effect of the cavity is to reduce the minimum detectable magnetic field value by a factor equal to the finesse $F_S\approx\pi/\varepsilon$. For $\varepsilon=\sqrt{a_{0}a_{\Omega_R}}$ (case ii), the contrast is equal to zero and $\delta B$ reaches a singular value as shown in Eq. (\ref{sensitivity}). Finally, for $\varepsilon<\sqrt{a_{0}a_{\Omega_R}}$ (case iii), assuming $\varepsilon\ll a_0$ the sensitivity reads
\begin{equation}\label{sensitivity_high_finesse}
\delta B\approx\frac{\Gamma_{\mathrm{mw}}a_{\Omega_R}a_{0}}{4\gamma\varepsilon(a_{\Omega_R}-a_0)}\sqrt{\frac{hc}{P_{0,S}t_{\mathrm{m}}\lambda_S}}.
\end{equation}
This shows that the sensitivity can be greatly impaired (\textit{i.e.} $\delta B$ increases) if the empty cavity finesse ($\pi/\varepsilon$) is larger than that of a critically coupled cavity given by $\pi/(2a_0)$. Moreover, Eqs. (\ref{sensitivity_low_finesse}) and (\ref{sensitivity_high_finesse}) show that if the off- and on-resonance loss values $a_0$ and $a_{\Omega_R}$ are too close, the sensitivity is also impaired.

As a conclusion, the level $\left|6\right\rangle$ is always partly populated due to the non ideal branching ratio to the dark singlet state ($k_{35}\neq0$). This results in absorption of the IR probe beam, even in the microwave-off state (\textit{i.e.} no resonant microwaves applied) and the implementation of a cavity will also increase this effect and reduce the detected IR photon number $I_{\mathrm{out},S}$. Thus, the cavity induces simultaneously an increase in the contrast $\mathcal{C}$ and a reduction of the detected photon number in the IR beam. Consequently, for a given single-pass absorption, the cavity finesse cannot be arbitrarily increased and the magnetic field sensitivity $\delta B$ reaches a minimum value intrinsically limited by NV$^-$ photophysical parameters and by diamond intrinsic IR optical losses. Those effects are quantitatively described in the next section where numerical results are reported.

\subsection{Numerical calculations}\label{monolitic}

The output fields $E_{\mathrm{out,i}}$ both for the pump and IR signal are deduced from the input and intracavity forward and backward propagating fields $f_i(z)$ and $b_i(z)$ described Fig. \ref{cavite}a) using the slowly varying envelope approximation. Note that the intracavity absorption (obtained by solving the six-level rate equations) depends nonlinearly on the intracavity intensity $I_i(z)$ and thus a numerical optimization routine on $f_i(L)$ must be used to deduce the reflected powers both at pump and signal wavelengths for the target values\cite{Danckaert91} of $I_{0,i}$ (see Appendix \ref{optim} for details on the calculation method).\\
\indent We consider two NV$^-$ center concentrations\cite{Kubo11, Acosta10} i) configuration 1: $n=4.4\times10^{23}~\mathrm{m}^{-3}$ and $T_2^*=390~\mathrm{ns}$ ii) configuration 2: $n=28\times10^{23}~\mathrm{m}^{-3}$ and $T_2^*=150~\mathrm{ns}$. For high NV$^-$-center density, single-pass absorption is high and the system is less sensitive to parasitic optical losses, but the electron spin dephasing time is shorter than for less low density samples. For each of these configurations we analyze: i) the effect of the diamond crystal sample thickness, ii) the effect of the input power, and iii) that of the Q-factor of the cavity. The Q-factors are defined by $Q_i=2n_dLF_i/\lambda_i$, ($i\in\{P,S\}$) $n_d=2.4$ being the diamond refractive index and where we recall (see Eq. (\ref{finesse_appendix}) in Appendix \ref{analy}) that the finesse $F_i$ is defined by 
\begin{equation}
F_i=\frac{\pi\sqrt{\rho_{\mathrm{in},i}A_i}}{1-\rho_{\mathrm{in},i}A_i}.
\end{equation}
with $A_i$ the single-pass round-trip transmission. Note that in the case of a resonant pump field, the cavity is designed in order to reach exactly the critical coupling $A_P=\rho_{\mathrm{in},P}$ which gives the maximal intracavity pump field enhancement and the optimal pump energy transfer to the NV$^-$ ensemble.

\begin{figure}
\includegraphics[width=9cm]{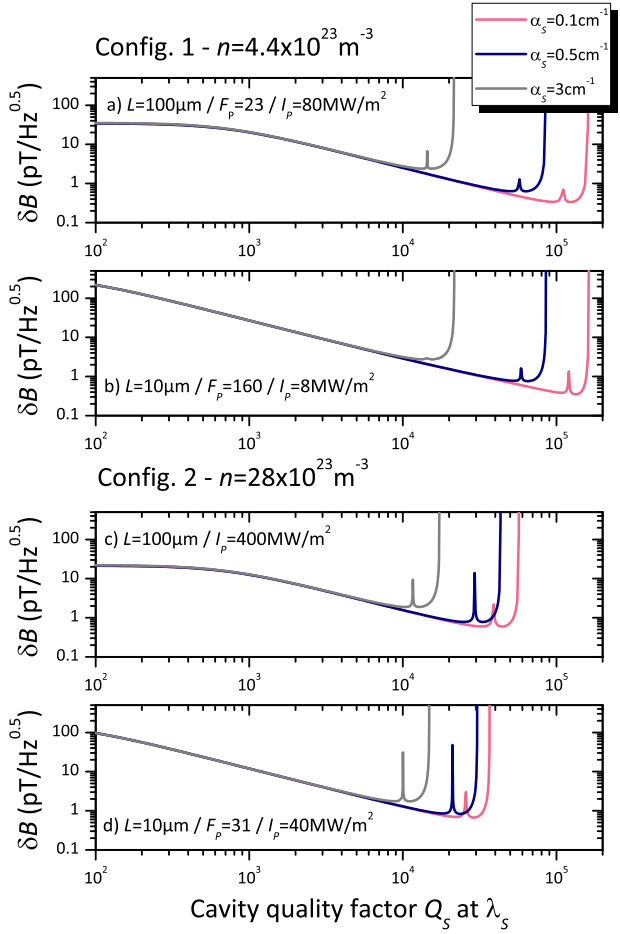}
\caption{Shot noise limited magnetic field sensitivity vs Q-factor of the cavity at the signal wavelength and for different values of IR-signal optical losses ($\alpha_S$). Calculations are done for $\Omega_R=2\pi\times1.5~\mathrm{MHz}$, $P_{0,S}=300~\mathrm{mW}$ with $I_{0,S}=150~\mathrm{MW/m}^2$ and no optical losses for the pump ($\alpha_P=0$). c) Config. 2 and $L=100~\mu\mathrm{m}$, we assume a single pass pumping. For each plot, the value of $\delta B$ obtained for low $Q_s$ is about half compared to that obtained for single-pass propagation as expected from the use of a high reflectivity backside mirror.\label{Finesse}}
\end{figure}

Figure \ref{Finesse} shows the magnetic-field sensitivity as a function of the cavity Q-factor $Q_S$ at the IR-signal wavelength for two cavity lengths and three values of $\alpha_S$ which represents the IR-signal optical-loss due to the bulk diamond material alone.
In the rate-equation approximation, the sensitivity reaches two maxima (minima of $\delta B$), the first corresponding to a cavity critically coupled when the microwaves are switched-on and the second corresponding to a cavity critically coupled when the microwaves are switched-off. Between these two optimal coupling configurations, we observe a sharp decrease of the sensitivity corresponding to a cancellation of the contrast. For this particular situation, the reflection for the microwave switched-on and switched-off cases are equal. The IR optical losses reduce the sensitivity of the cavity but for $\alpha_S=0.5~\mathrm{cm}^{-1}$ ($\alpha_S=0.1~\mathrm{cm}^{-1}$) the best sensitivity can reach $0.6~\mathrm{pT}/\sqrt{\mathrm{Hz}}$ ($0.3~\mathrm{pT}/\sqrt{\mathrm{Hz}}$) corresponding to almost two orders of magnitude enhancement in comparison to single-pass approaches. For strong optical losses ($\alpha_S=3~\mathrm{cm}^{-1}$) the sensitivity is still enhanced by more than one order of magnitude and the performance of the cavity system is comparable with that of the same sample in a single-pass configuration at low temperature \cite{Acosta10}.\\
\indent We now discuss the results for IR optical losses set to $\alpha_S=0.5~\mathrm{cm}^{-1}$. For $n=4.4\times10^{23}~\mathrm{m}^{-3}$, it is possible to use a doubly resonant cavity to increase the intracavity optical pump intensity and thus to reduce the required external intensity as illustrated in Fig. \ref{cavite}c). By diminishing the length of the cavity, the single pass attenuation is reduced and thus it is possible to increase the pump cavity finesse and thus to strongly reduce the required amount of pump power from $400~\mathrm{MW/m}^2$ (single-pass propagation) to $8~\mathrm{MW/m}^2$. For $n=28\times10^{23}~\mathrm{m}^{-3}$, the pump absorption is so high that for $L=100~\mu\mathrm{m}$ a doubly resonant approach does not give any improvement in the required pump power ($I_{0,P}=400~\mathrm{MW/m}^2$). Nevertheless, for short cavities ($L=10~\mu\mathrm{m}$) a modest-finesse cavity for the pump ($F_P=31$) leads to a reduction of the external pump power (down to $I_{0,P}=40~\mathrm{MW/m}^2$).
\begin{figure}
\includegraphics[width=9.25cm]{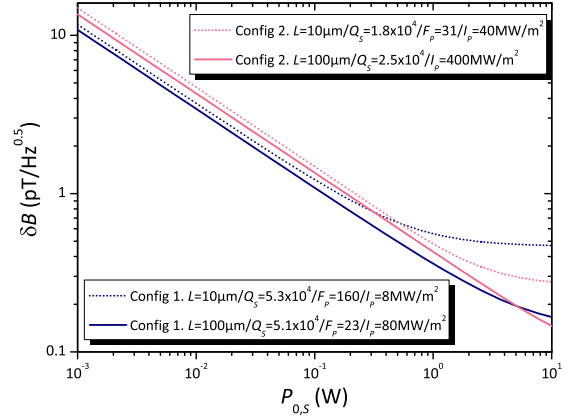}
\caption{Shot-noise limited magnetic-field sensitivity calculated for $\Omega_R=2\pi\times1.5~\mathrm{MHz}$, $\alpha_S=0.5~\mathrm{cm}^{-1}$, $\alpha_P=0$ and $2w_0=50~\mu\mathrm{m}$ varying the input IR signal power. For Config.~2 ($n=28\times 10^{23}~\mathrm{m}^{-3}$) and $L=100~\mu\mathrm{m}$ we assume single-pass propagation for the pump. The cavity parameters have been optimized using Fig. \ref{Finesse}. \label{Saturation}}
\end{figure}
In Fig. \ref{Saturation} we plot the magnetic field sensitivity as a function of the IR signal input power $P_{0,S}$ for a beam-waist diameter $2w_0=50~\mu\mathrm{m}$. For thick diamond slabs, the saturation is obtained at high power ($\geq10~\mathrm{W}$). For thin diamond slabs, the use of high-finesse cavities reduces the signal saturation power. In the highest-Q-factor case (Config.~1 and $L=10~\mu\mathrm{m}$), saturation starts around $P_{0,S}\approx300~\mathrm{mW}$. For high signal input power thermal effects must be taken into account. Note that these effects would improve the sensitivity via the thermo-optic effects. More generally any nonlinear dispersive effect would increase the sensitivity of the device. In this case, a change in the absorption for the signal would induce a shift of the cavity resonance. In the example of Fig. \ref{cavite}.b), if we denote $\lambda_1-\lambda_0$ the shift of the cavity, the contrast would be given by $[R_S(0,\lambda_1)-R_S(\Omega_R)]/R_S(0,\lambda_1)$ and would have approximately the same value than without nonlinear effects. However the detected reflected power would be $R_S(0,\lambda_1)\times P_{0,S}$ and would be greatly increased in comparison with $R_S(0)\times P_{0,S}$ which could reduce the value of the minimum of the detectable magnetic field as shown for example by Eq. (\ref{sensitivity}).\\
\indent We can check that all the results given here are consistent with the quantum-noise limited sensitivity: i) Config. 1 $\delta B_{\mathrm{q}}=0.2~\mathrm{pT}/\sqrt{\mathrm{Hz}}$ and $\delta B_{\mathrm{q}}=0.06~\mathrm{pT}/\sqrt{\mathrm{Hz}}$ ii) Config 2. $\delta B_{\mathrm{q}}=0.13~\mathrm{pT}/\sqrt{\mathrm{Hz}}$ and $\delta B_{\mathrm{q}}=0.04~\mathrm{pT}/\sqrt{\mathrm{Hz}}$ for $L=10~\mu\mathrm{m}$ and $L=100~\mu\mathrm{m}$ respectively. The choice of parameters for each case considered above results from an optimization depending on the crystal thickness and NV$^-$ center concentration. Note that in the most resonant configuration (Config.~1 and $L=10~\mu\mathrm{m}$), the optimal overall Q-factor of the cavity for the probe is around $5.3\times10^4$, giving a cavity bandwidth $\gamma_{\mathrm{cav}}=2\pi\times5.4~\mathrm{GHz}$ much larger than the probe-laser linewidth ($\gamma_{L}\approx2\pi\times10~\mathrm{MHz}$) used for single-pass experiments reported in Ref. [\onlinecite{Acosta10}]. For high NV$^-$ concentrations (Config.~2), the required Q-factor can be low ($\leq3\times10^4$) and thus the total optical path $\ell=\lambda_SQ_S/(2\pi n_d)$ ($\ell\approx2~\mathrm{mm}$) is smaller or almost equal to the Rayleigh range obtained for a waist diameter $2w_0=50~\mu\mathrm{m}$ ($2Z_R\approx3.8~\mathrm{mm}$). Consequently, the simple planar Fabry-Perot geometry\cite{Dumeige11} depicted in Fig \ref{cavite}a) can be used. Finally, considering highly concentrated thin samples the required Q-factor can be around $2\times10^4$ which is compatible with recent measurement reported on integrated diamond microcavities\cite{Hausmann12}. 

\subsection{External-mirror cavities}

For the highest-finesse cavities, appropriate for a concentration of $n=4.4\times10^{23}~\mathrm{m}^{-3}$, the effective length $\ell$ is longer than the Rayleigh range for the chosen beam waist value ($2w_0=50~\mu\mathrm{m}$). Consequently, external spherical mirrors should be used. If we consider for example a confocal cavity, the distance between the mirrors is $L_{\mathrm{cav}}=2Z_R=3.8~\mathrm{mm}$. For a $100~\mu\mathrm{m}$ ($10~\mu\mathrm{m}$) thick diamond plate, the finesse of the cavity would be $F_S=110$ ($F_S=1150$). Consequently, in the case of the highest finesse cavity, the Q-factor would be $8.4\times10^6$ corresponding to a cavity bandwidth $\gamma_{\mathrm{cav}}=2\pi\times34~\mathrm{MHz}$ still larger than the probe-laser linewidth. We have assumed here distributed optical losses such as $\alpha_S=0.5~\mathrm{cm}^{-1}$; if we consider that optical losses mainly come from diamond interface roughness, it implies that in the more unfavorable case (for the $10~\mu\mathrm{m}$-thick diamond plate), the root mean square deviation of the surface to planarity of the diamond interfaces\cite{Lerondel99} has to be less than $2~\mathrm{nm}$, which is attainable with state-of-the-art fabrication techniques\cite{Koslowski00}. 

\section{Conclusion}

The use of a cavity can enhance the sensitivity of optical magnetometers based on IR absorption of NV$^-$ centers in diamond at room temperature. We found that for diamond samples with a high density of defects (NV$^-$-center concentration larger than $n\geq4.4\times10^{23}~\mathrm{m}^{-3}$), our configuration allows an enhancement of two orders of magnitude in comparison with single-pass configurations. In the presence of high IR optical losses the enhancement is reduced to one order of magnitude. The use of a cavity compensates for the reduction of the optical depth due to homogeneous broadening at room temperature \cite{Acosta10}. Moreover, doubly resonant (for the pump and the probe) cavities can be used to reduce the amount of required pump intensity (down to $8~\mathrm{MW/m}^2$). Using diamond samples with a very high density of defects ($n\approx28\times10^{23}~\mathrm{m}^{-3}$), this approach could be implemented using monolithic planar Fabry-Perot cavities or integrated diamond photonic structures such as microdisk or microring resonators. For smaller defect concentrations ($n\approx4.4\times10^{23}~\mathrm{m}^{-3}$), external spherical-mirror cavities should be used.

\begin{acknowledgments}
This work was supported by the France-Berkeley Foundation, the AFOSR/DARPA QuASAR program, IMOD and the NATO Science for Peace program. K.J. was supported by the Danish Council for Independent Research Natural Sciences.
\end{acknowledgments}

\appendix

\section{NV$^-$ six-level modeling}\label{rate}

The local density $n_j(z)$ (with $j\in\left[1,6\right]$) of the centers of each level are calculated by solving the rate equations assuming $dn_j/dt=0$. We consider spin-conserving optical transitions. The pump excites a vibronic sideband which decays quickly via phonon emission to levels $\left|3\right\rangle$ and $\left|4\right\rangle$. This allows us to neglect the down-transition rates due to the pump light. At $z$, the relation between the optical intensity and the center densities is given by
\begin{equation}\label{calcul_N}
\mathcal{M}(z)\cdot\mathcal{N}(z)=\mathcal{N}_0,
\end{equation}
where $\mathcal{N}_0=(0,0,0,0,0,n)^T$, $\mathcal{N}$ contains the values of the center densities: $\mathcal{N}=(n_1,n_2,n_3,n_4,n_5,n_6)^T$ and the matrix $\mathcal{M}(z)$ can be written:
\begin{widetext}
\begin{equation}
  \mathcal{M}(z)=\left(
    \begin{array}{c c c c c c}
      -[W_P(z)+W_{\mathrm{mw}}] & W_{\mathrm{mw}} & k_{31} & 0 & 0 & k_{61}\\
       W_{\mathrm{mw}} & -[W_P(z)+W_{\mathrm{mw}}] & 0 & k_{42} & 0 & k_{62}\\
       W_P(z) & 0 & -(k_{31}+k_{35}) & 0 & 0 & 0\\
       0 & W_P(z) & 0 & -(k_{42}+k_{45}) & 0 & 0\\
       0 & 0 & k_{35} & k_{45} & -[W_S(z)+\Gamma] & +W_S(z)\\
      1 & 1 & 1 & 1 & 1 & 1\\
    \end{array}
  \right),
\end{equation}
\end{widetext}
we assume here a closed system: $\sum_{j=1}^6n_j=n$. The transition rates $W_i$ ($i=P$ for the pump and $i=S$ for the IR signal) are related to the optical intensity $I_i$, the wavelength $\lambda_i$ and the absorption cross section $\sigma_i$ by $W_i=\sigma_iI_i\lambda_i/(hc)$. Assuming a low Rabi angular frequency $\Omega_R$, in the rate-equation approximation, the microwave transition rate is calculated as $W_{\mathrm{mw}}=\Omega_R^2T^*_2/2$ where $T^*_2$ is the electron spin dephasing time. The center density in each level is calculated by $\mathcal{N}=\mathcal{M}^{-1}\mathcal{N}_0$.

\section{IR-absorption cross section estimation}\label{cross_section_estimation}

In order to model the system we have to evaluate the IR absorption (due to singlet states) cross section $\sigma_S$ which has not been measured so far. With the aim of designing a cavity based magnetometer, the value of $\sigma_S$ is important to evaluate the intracavity IR signal intensity saturation. This completes the already reported list of photophysical properties of the NV$^-$ centers in diamonds that are summarized in Tab. \ref{tab_nv} given in Appendix \ref{tableaux}. Here we estimate $\sigma_S$ by using the single-pass IR-absorption measurements described in Ref.~[\onlinecite{Acosta10}]. We assume that the measured magnetic field is oriented in such a way that the microwaves are only resonant with NV$^-$ centers of a particular orientation, \textit{i.e.}, one quarter of all the NV$^-$ centers\cite{Pham12}. In the single-pass configuration $\mathcal{C}$ can be calculated by integrating the two differential equations considering off-resonance pumping and a resonant excitation (including stimulated emission) for the signal
\begin{equation}
  \left\{
    \begin{aligned}
      \frac{dI_P}{dz} & = -\left\{\sigma_P\left[n_1(z)+n_2(z)\right]+\alpha_P\right\}I_P(z)\\
\frac{dI_S}{dz} & = -\left\{\sigma_S\left[n_6(z)-n_5(z)\right]+\alpha_S\right\}I_S(z),\\
    \end{aligned}
  \right.
\end{equation}
where the densities $n_i(z)$ with $i\in[1,6]$ are the stationary solutions of the rate equations corresponding to Fig. \ref{level}a) (see Appendix \ref{rate}). $\alpha_i$ with $i\in\{P,S\}$ are the optical losses due to light scattering or parasitic absorption. Calculations are carried out using the parameters given in Ref. [\onlinecite{Acosta10}] recalled in Tab. \ref{tab_nv2} (see Appendix \ref{tableaux}). The two unknown values are the IR absorption cross section $\sigma_S$ and the optical losses $\alpha_P$ at the pump wavelength. The method consists in numerically finding the values of $\sigma_S$ which gives the contrast value defined in Eq. (\ref{Contrast_single_pass}) and reported in Ref. [\onlinecite{Acosta10}]. We have then deduced that for a monochromatic excitation (the linewidth of the IR laser is $\gamma_L\approx 2\pi\times10~\mathrm{MHz}\ll\gamma_{\mathrm{IR}}$), the IR absorption cross section due to the metastable level is $\sigma_S=(2.0\pm0.3)\times10^{-22}~\mathrm{m}^2$. The uncertainties come from the value of $\alpha_P$ which has been assumed to vary from $0$ to $10~\mathrm{cm}^{-1}$. The associated saturation intensity is $I_{\mathrm{sat},S}=hc\Gamma/(2\lambda_S\sigma_S)\approx500~\mathrm{GW/m}^2$.

\section{Analytic expression of the cavity-reflectivity in the linear regime}\label{analy}

Here we consider the cavity described in Fig. \ref{cavite}a) with $\rho_{\mathrm{back},S}=1$. We denote the probe input field $E_{0,S}$, the reflected field $E_{\mathrm{out},S}$ and the forward propagating field inside the cavity at the input mirror $\mathcal{F}_S(0)$. Introducing the amplitude mirror IR transmission coefficient $\kappa_{\mathrm{in},S}$ verifying $\kappa_{\mathrm{in},S}^2+\rho^2_{\mathrm{in},S}=1$ and the round-trip phase $\varphi$, we can write
\begin{equation}
  \left\{
    \begin{aligned}
      \mathcal{F}_S(0) & =j\kappa_{\mathrm{in},S}E_{0,S}+\rho_{\mathrm{in},S}A_S\mathcal{F}_S(0)e^{j\varphi}\\
      E_{\mathrm{out},S} & =\rho_{\mathrm{in},S}E_{0,S}+j\kappa_{\mathrm{in},S}A_S\mathcal{F}_S(0)e^{j\varphi}.\\
    \end{aligned}
  \right.
\end{equation}
By eliminating $\mathcal{F}_S(0)$, we can deduce the amplitude transfer function of the cavity
\begin{equation}
\frac{E_{\mathrm{out},S}}{E_{0,S}}=\frac{\rho_{\mathrm{in},S}-A_Se^{j\varphi}}{1-\rho_{\mathrm{in},S}A_Se^{j\varphi}}.
\end{equation}
The intensity reflectivity of the cavity is thus given by
\begin{equation}
\left|\frac{E_{\mathrm{out},S}}{E_{0,S}}\right|^2=\frac{\rho_{\mathrm{in},S}^2+A_S^2-2\rho_{\mathrm{in},S}A_S\cos{\varphi}}{1+\rho_{\mathrm{in},S}^2A_S^2-2\rho_{\mathrm{in},S}A_S\cos{\varphi}}.
\end{equation}
At resonance $\varphi=0~(2\pi)$ the reflectivity of the cavity can be written
\begin{equation}
R_S=\left(\frac{\rho_{\mathrm{in},S}-A_S}{1-\rho_{\mathrm{in},S}A_S}\right)^2.
\end{equation}
In the all-pass configuration, the finesse of the cavity is given by
\begin{equation}\label{finesse_appendix}
F_S=\frac{\pi\sqrt{\rho_{\mathrm{in},S}A_S}}{1-\rho_{\mathrm{in},S}A_S}.
\end{equation}

\section{Numerical cavity-reflectivity calculation}\label{optim}

For $i\in\{S,P\}$, if $\mathcal{F}_i$ and $\mathcal{B}_i$ denote the forward and backward propagating fields, the intracavity field $E_i$ can be written
\begin{equation}
E_i(z)=\mathcal{F}_i(z)+\mathcal{B}_i(z).
\end{equation}
With $f_i$ and $b_i$, the slowly varying envelope amplitudes of the forward and backward propagating fields shown in Fig. \ref{cavite}a), we obtain 
\begin{equation}
E_i(z)=f_i(z)e^{-j\beta_iz}+b_i(z)e^{j\beta_iz},
\end{equation}
with $\beta_i=2\pi n_d/\lambda_i$. The field amplitudes are normalized in order to have $I_i(z)=\left|E_i(z)\right|^2$. The calculation of the cavity reflection is a two point boundary value problem. It can be solved by a shooting method. The first boundary condition is that there is no incoming field from the $z>0$. This can be written by the following relation between the forward and backward propagating field values at the back mirror
\begin{equation}
b_i(L)=\rho_{\mathrm{back},i}f_i(L)e^{-2j\beta_iL}.
\end{equation}
From this starting values we can deduce the values of the envelope amplitudes at the input mirror by integrating the following differential coupled equations  
\begin{equation}
  \left\{
    \begin{aligned}
      \frac{df_P}{dz} & =-\frac{1}{2}\left\{\sigma_P[n_1(z)+n_2(z)]+\alpha_P\right\}f_P(z)\\
      \frac{db_P}{dz} & =\frac{1}{2}\left\{\sigma_P[n_1(z)+n_2(z)]+\alpha_P\right\}b_P(z)\\
      \frac{df_S}{dz} & =-\frac{1}{2}\left\{\sigma_S[n_6(z)-n_5(z)]+\alpha_S\right\}f_S(z)\\
      \frac{db_S}{dz} & =\frac{1}{2}\left\{\sigma_S[n_6(z)-n_5(z)]+\alpha_S\right\}b_S(z),\\
    \end{aligned}
  \right.
\end{equation}
where the values of the NV$^-$ center density are deduced from Eq. (\ref{calcul_N}). We can obtain the input $I_{0,i}=\left|E_{0,i}\right|^2$ and output $I_{\mathrm{out},i}=\left|E_{\mathrm{out},i}\right|^2$ intensities from
\begin{equation}
  \left\{
    \begin{aligned}
      E_{0,i} & =\frac{1}{j\kappa_{\mathrm{in},i}}\left[f_i(0)-\rho_{\mathrm{in},i}b_i(0)\right]\\
      E_{\mathrm{out},i} & =\rho_{\mathrm{in},i}E_{0,i}+j\kappa_{\mathrm{in},i}b_i(0),\\
    \end{aligned}
  \right.
\end{equation}
where $\kappa_{\mathrm{in},i}$ for $i\in\{P,S\}$ ($\kappa_{\mathrm{in},i}^2+\rho^2_{\mathrm{in},i}=1$) are the amplitude mirror transmission coefficients. The calculation method consists in numerically optimizing the values of $f_i(L)$ to obtain the target values of $I_{0,i}$. The value of $R_S=I_{\mathrm{out},S}/I_{0,S}$ is then deduced with and without the microwave field applied. This is used to calculate the contrast $\mathcal{C}$ using Eq. (\ref{contrast_cav}) and the effective detected power $\mathrm{max}[R_S(\Omega_R),R_S(0)]\times P_S$. Finally, the minimum detectable magnetic field $\delta B$ is evaluated using Eq. (\ref{sensitivity_cav}).

\section{Tables}\label{tableaux}

\begin{table}[h]
\renewcommand*\arraystretch{1.3}
\caption{\label{tab_nv}Photophysical parameters of the six-level system sketched in Fig. \ref{level}a). The transition rates $k_{ij}$ are obtained by averaging data given in Ref. [\onlinecite{Tetienne12}]. $1/\Gamma$ is the lifetime of level $\left|5\right\rangle$. $\gamma_{\mathrm{IR}}$ is the spectral width of the $1042~\mathrm{nm}$ zero-phonon line at room temperature.}
\begin{tabular}{c | c | c}
 Parameter & Value & Reference\\
 \hline
  $\lambda_P$        &    $532~\mathrm{nm}$               & [\onlinecite{Acosta10}]\\
  $\lambda_S$        &    $1042~\mathrm{nm}$              & [\onlinecite{Acosta10}]\\
	$\sigma_P$         &    $3\times10^{-21}~\mathrm{m}^2$  & [\onlinecite{Wee07}]\\
	$k_{31}=k_{42}$    &    $(66\pm5)~\mu\mathrm{s}^{-1}$     & [\onlinecite{Tetienne12}]\\
	$k_{35}$           &    $(7.9\pm4.1)~\mu\mathrm{s}^{-1}$      & [\onlinecite{Tetienne12}]\\
	$k_{45}$           &    $(53\pm7)~\mu\mathrm{s}^{-1}$      & [\onlinecite{Tetienne12}]\\
	$k_{61}$           &    $(1.0\pm0.8)~\mu\mathrm{s}^{-1}$      & [\onlinecite{Tetienne12}]\\
	$k_{62}$           &    $(0.7\pm0.5)~\mu\mathrm{s}^{-1}$      & [\onlinecite{Tetienne12}]\\
	$\Gamma$           &    $1~\mathrm{ns}^{-1}$            & [\onlinecite{Acosta10prb}]\\
	$\gamma_{\mathrm{IR}}$      &    $\approx2\pi\times4~\mathrm{THz}$                & \\
\end{tabular}
\end{table}

\begin{table}[h]
\renewcommand*\arraystretch{1.3}
\caption{\label{tab_nv2}Physical parameters used in the single-pass NV$^-$ center IR absorption measurements\cite{Acosta10} at room temperature. Optical losses are estimated from the transmission spectrum given in Ref. [\onlinecite{Acosta09}].}
\begin{tabular}{c | c | c}
 Parameter & Value & Reference\\
 \hline
	$n$            &        $28\times10^{23}~\mathrm{m}^{-3}$     & [\onlinecite{Acosta10}]\\
	$T^*_2$        &        $150~\mathrm{ns}$                    & [\onlinecite{Acosta10}]\\
	$T_1$        &          $2.9~\mathrm{ms}$                    & [\onlinecite{Jarmola2012}] ($\mathrm{T}=300~\mathrm{K}$)\\
	$I_{0,P}$      &        $400~\mathrm{MW}/\mathrm{m}^2$       & [\onlinecite{Acosta10}]\\
	$I_{0,S}$      &        $10~\mathrm{MW}/\mathrm{m}^2$        & \\
	$P_{S}$        &        $16~\mathrm{mW}$                     & \\
	$L$            &        $300~\mu\mathrm{m}$                  & [\onlinecite{Acosta10}]\\
	$\Omega_R$     &        $2\pi\times1.5~\mathrm{MHz}$                   & [\onlinecite{Acosta10}]\\
	$\mathcal{C}$  &        $0.003$                              & [\onlinecite{Acosta10}] ($\mathrm{T}=300~\mathrm{K}$)\\
	$\alpha_S$     &        $0.1-0.5~\mathrm{cm}^{-1}$               & [\onlinecite{Acosta09}]\\
\end{tabular}
\end{table}
The large value of $T_1$ shows that spin relaxation is negligible. Thus this is not taken into account in the rate-equation modeling of the six-level system.

\section{Sensitivity fundamental limit}\label{fundamental}

In this Appendix we derive the fundamental limit of the minimal detectable magnetic field value for methods based on IR absorption or visible fluorescence monitoring considering that the methods are limited by the photon shot-noise.

\subsection{IR absorption based magnetometer}

Using the expression of the ESR FWHM and Eq. (\ref{sensitivity}) we obtain the following relation for the minimal detectable magnetic field
\begin{equation}
\delta B=\frac{2}{\gamma\mathcal{C}\sqrt{N_{\mathrm{ph}}\left(T_2^*\right)^2t_{\mathrm{m}}}}
\end{equation}
where $N_{\mathrm{ph}}=P_S\lambda_S/(hc)$ is the number of detected IR photons per second. With $N_S$ the number of IR photons collected per $T_2^*$ we have
\begin{equation}
\delta B=\frac{2}{\gamma\mathcal{C}\sqrt{N_{S}T_2^*t_{\mathrm{m}}}}.
\end{equation}
Now we estimate the maximal $N_S$ value. Assuming an optimal contrast $\mathcal{C}=1$. When microwaves are switched-on, every photon is absorbed. We assume that one NV$^-$ center absorbs $M_S$ IR photons per $T_2^*$. In many high-density samples, $T_2^*\lesssim1/(k_{61}+k_{62})$, and therefore we can consider that $M_S<\Gamma T_2^*$. We can thus write
\begin{equation}
N_S=M_S\left(N_{\mathrm{on}}^{\mathrm{sing}}-N_{\mathrm{off}}^{\mathrm{sing}}\right),
\end{equation}
where $N_{\mathrm{on}}^{\mathrm{sing}}$ is the number of NV$^-$ centers in the singlet state when the microwave are switched-on and $N_{\mathrm{off}}^{\mathrm{sing}}$ the number of NV$^-$ centers in the singlet for switched-off microwaves. This gives the number of photons which can be detected when the microwaves are switched-off
\begin{equation}
N_S=M_SN\left[\left(\frac{3}{4}\times P_{35}+\frac{1}{4}\times P_{45}\right)-P_{35}\right],
\end{equation}
where $N=nV$ is the number of centers with $P_{35}=k_{35}/(k_{35}+k_{31})$ being the probability that NV$^-$ centers in level $\left|3\right\rangle$ ($m_s=0$) decay to the singlet and $P_{45}=k_{45}/(k_{45}+k_{42})$ the probability that NV$^-$ centers in level $\left|4\right\rangle$ ($m_s=\pm 1$) decay to the singlet. The $\frac{1}{4}$ and $\frac{3}{4}$ allow to take into account that only one quarter of the NV$^-$ centers are resonant with the microwaves\cite{Pham12}. We then have $N_S=\mathcal{R}_SM_SN$ with
\begin{equation}
\mathcal{R}_S=\frac{1}{4}\left(\frac{k_{45}}{k_{45}+k_{42}}-\frac{k_{35}}{k_{35}+k_{31}}\right),
\end{equation}
which is an approximated value for $R_S(\Omega_R)$ defined in section \ref{monolitic}. Note that if the IR power is such as $\mathcal{R}_SM_S\geq1$ the sensitivity is limited by the spin-noise.

\subsection{Fluorescence measurement based magnetometer}

For a magnetometer using the fluorescence signal monitoring and assuming that the ESR FWHM is $2/T_2^*$, the sensitivity is given by\cite{Dreau11, Rondin12}
\begin{equation}
\delta B_{\mathrm{f}}=\frac{2}{\gamma\mathcal{C}_{\mathrm{f}}\sqrt{N_{\mathrm{f}}T_2^*t_{\mathrm{m}}}},
\end{equation}
where $\mathcal{C}_{\mathrm{f}}$ is the contrast of the fluorescence signal and $N_{\mathrm{f}}$ the number of collected photons per $T_2^*$. When the microwaves are switched-off, the fluorescence signal is proportional to $P_{31}=k_{31}/(k_{31}+k_{35})$ the probability that NV$^-$ centers in level $\left|3\right\rangle$ decay immediately to level $\left|1\right\rangle$. When the microwaves are switched-on the fluorescence signal is proportional to $P_{42}/4+3P_{31}/4$ where $P_{42}=k_{42}/(k_{42}+k_{45})$ is the probability that NV$^-$ centers in level $\left|4\right\rangle$ decay to level $\left|2\right\rangle$. Assuming that $P_{31}\approx1$ ($k_{35}\ll k_{31}$), the contrast $\mathcal{C}_{\mathrm{f}}$ is given by
\begin{equation}
\mathcal{C}_{\mathrm{f}}=\frac{1}{4}\left(\frac{k_{31}}{k_{31}+k_{35}}-\frac{k_{42}}{k_{42}+k_{45}}\right).
\end{equation}
The number of collected photons per $T_2^*$ is $N_{\mathrm{f}}=\eta N M_{\mathrm{f}}$ where $\eta$ is the collection efficiency and $M_{f}$ the number of emitted photons per $T_2^*$ by one NV$^-$ center. Since $1/k_{35}<T_2^*$ we have $M_{\mathrm{f}}<k_{31}/k_{35}$.

\subsection{Comparison}

\noindent The two techniques can be compared by calculating
\begin{equation}
\frac{\delta B_{\mathrm{f}}}{\delta B}\approx\frac{1}{\mathcal{C}_{\mathrm{f}}}\sqrt{\frac{\mathcal{R}_SM_S}{\eta M_{\mathrm{f}}}},
\end{equation}
where we assume $\mathcal{C}\approx1$. For $k_{35}\ll k_{31}$ and $k_{42}\approx k_{45}$, we have $\mathcal{R}_S\approx\mathcal{C}_{\mathrm{f}}$ and thus
\begin{equation}
\frac{\delta B_{\mathrm{f}}}{\delta B}\approx\sqrt{\frac{M_S}{M_{\mathrm{f}}}\cdot\frac{1}{\eta\mathcal{R}_{S}}}.
\end{equation}
Note that with values recalled in Tab. \ref{tab_nv2}, we obtain $\mathcal{R}_{S}\approx8.5\%$ which corresponds to the optimal case asuming a total spin polarization. We deduce that $M_S\leq11$ and $M_\mathrm{f}\leq8$. Assuming that $M_S=M_\mathrm{f}$ and considering a high value of the collection efficiceny ($\eta\approx0.47$ has been reported in Ref. [\onlinecite{LeSage2012}]) we obtain $\delta B_\mathrm{f}/\delta B\approx5$.

\bibliography{NV_biblio}

\end{document}